# YANK AND HOOKE'S CONSTANT GROUP THEORETICALLY

*Joachim Nzotungicimpaye*[(*)]

*Kigali Institute of Education ,*

*P.O.Box 5039, Kigali, Rwanda*

*e-mail address* kimpaye@yahoo.fr

**Abstract**. *We study the second central extension* $G_2$ *of the (1+1) Aristotle Lie group* $G_0$. *We find that the first central extension* $G_1$ *of* $G_0$ *admits four two dimensional orbits on the dual* $G^*_2$ *of the Lie algebra of* $G_2$. *The generic orbit is characterised by a Hooke's constant k and a yank y. If the physics of the orbit is studied with respect the evolution in time , it represents an elementary system with internal energy* U *in a position-momentum (q,p) state under the conjugation of a Hooke's force kq and a damping force* $c(t)\frac{dq}{dt}$, $\frac{dq}{dt}$ *being a velocity as in particle mechanics*. *If the physics of the orbit is studied with respect the evolution in space , it represents an elementary system with internal linear momentum P in a energy-time (e, $\tau$ ) state under the conjugation of a kind of Hooke's force $y\tau$ and a damping force* $W(x)\frac{d\tau}{dx}, \frac{d\tau}{dx}$ *being a slowness as in time travel waves*.

1.**Introduction**

It is well known that the first and the second derivative of position are respectively called velocity and acceleration. It is less well known that the third derivative is technically known as jerk ([3],[6]).As its name suggests jerk is applied when evaluating the destructive effect of motion on a mechanism or discomfort caused to passengers in a vehicle. It has been also suggested that the fourth derivative be called snap [6].

---

*(*) On leave from absence of Faculty of Sciences, University of Burundi, Bujumbura, Burundi*

Moreover we know that force is rate of change of momentum. It has been proposed [6] that yank and tug be respectively the rate of change of force and the rate of change of yank. We can then say with [6]:

"Momentum equals mass times velocity;

Force equals mass times acceleration,

Yank equals mass times jerk,

Tug equals mass times snap".

Recently we introduced jerk group theoretically [3] by symplectic realization of the first central extension of the (1+1) Galilei Lie group. Note that this group can be seen as a first central extension of the velocity-time translation Lie group .For this reason , the mass and the force found in ([3],[4]) can be considered as dual to the parameters extending the velocity-time translation Lie group for the second time while jerk is dual to a parameter within those extending the velocity-time translation Lie group for the third time.

In this paper, we study the second central extension of the (1+1) Aristotle Lie group [1] which is the(1+1) space-time translation Lie group. We show that yank and Hooke's constant come as dual to the extending parameters in the second extension. This can be compared mass and force in the (1+1) the velocity-time translation Lie group.

The mathematical expositions in this paper are not new because the (1+1) Aristotle Lie group is mathematically isomorphic to the velocity-time translation group whose the first central extension is exactly the (1+1) Galilei group studied in [3] and [4].

What is new and different from the papers [3] and [4] is the physics of the orbits. We find group theoretically new physical quantities such as Hooke's constant analogue to mass, yank analogue to force. It is these physical novelties which justify this paper. It is organized as follows. In section 2, we classify the co-adjoint orbits of the central extension of the (1+1) Aristotle Lie group on the dual of the second central extension Lie algebra. In the section 3,we study in details the physics of the generic orbit first when evolution with time is considered and then when evolution with space is considered. In the first case we find that the orbit is an elementary system under which act a conjugation of a Hooke's force and a damping one [7] while it a time travel wave with a slwoness $s$ [10] is in the second case. Yank is to elapsed time what the Hooke's constant is to elongation (travelled space) and power is to slowness what the damping coefficient is to velocity .

2. **Second central extension of the Aristotle Lie group**

It is known [1] that the (1+1) Aristotle Lie group $G_o$ is an additive Lie group whose the parameters are the space $x$ and time translations $t$. It is an additive abelian one , parametrized by space translation $x$ and the time translations $t$.

Using the standard methods [2] we can verify that the central extension $G_1$ of the Lie algebra $G_o$ of $G_o$ is generated by $P$ (for space translations),$E$ (for time translations) and $F$ (extending generator) , such that the nontrivial Lie bracket is

$$[P,E] = F \qquad (2.1)$$

If we write the general element of the connected Lie group $G_1$ associated to $\mathcal{G}_1$ as

$$g_1 = \exp(\varsigma F)\exp(xP + tE) \quad (2.2)$$

and if we use of the Baker-Campbell-Hausdorff formulae and the notion of trivial cocycle [3], we then find that the multiplication law for the group $G_1$ is

$$(x,t,\varsigma)(x',t',\varsigma') = (x+x', t+t', \varsigma+\varsigma'+xt') \quad (2.3)$$

Note that $G_1$ is mathematically isomorphic to the (1+1) Galilei group ([3],[4]) However the study of the co-adjoint orbits of $G_1$ on the dual of his extended Lie algebra (see next section) show that it has a different physical meaning.

Using again the standard methods we verify ([3],[4] and references therein) that the second central extension Lie algebra $\mathcal{G}_2$ of the Aristotle Lie algebra is generated by $P, E, F, \Lambda, Y$ such that the nontrivial Lie brackets are

$$[P,E] = F, [P,F] = \Lambda, [F,E] = Y \quad (2.4)$$

The general element of the corresponding connected Lie group $G_2$ can be written as

$$g_2 = \exp(a\Lambda + bY)\exp(tE + \varsigma F)\exp(xK) \quad (2.5)$$

By use of the Baker-Campbell-Hausdorff formulae we can verify that the resulting multiplication law is

$$(x,t,\varsigma,a,b)(x',t',\varsigma',a',b')$$
$$= (x+x', t+t', \varsigma+\varsigma'+xt', a+a'+x\varsigma'+\frac{1}{2}x^2 t', b+b'+\varsigma t'+\frac{1}{2}xt'^2) \quad (2.6)$$

Let $X = (\delta v, \delta t, \delta x, \delta\xi, \delta\varsigma)$ and $\mu = (p, e, f, k, y)$ be respectively an infinitesimal displacement on $G_2$ and an element of the dual $G^*_2$ of the Lie algebra $\mathcal{G}_2$ such that

$$\langle \mu, X \rangle = p\delta x + e\delta t + f\delta\varsigma + k\delta a + y\delb \quad (2.7)$$

is the associated scalar whose physical dimension is action.

We then verify that the co-adjoint action of $G_1$ on $G^*_2$ is

$$Ad^*_{(x,t,\varsigma)}(p,e,f,k,y)$$
$$= (p+ft+k(\varsigma-xt)+y\frac{t^2}{2}, e-fx+k\frac{x^2}{2}-y\varsigma, f-kx+yt, k, y) \quad (2.8)$$

As $\langle \mu, X \rangle$ has the dimension of an action, $e$, $p$, $f$, $k$ and $y$ have respectively the dimension of an energy, a linear momentum, a force, a spring resistance (Hooke constant) and a yank [6]. The last one being the rate of change of force with time.

We see from (2.8) that $k$ and $y$ are Ad*-invariants. As for the Galilei group [4], $G_1$ has four orbits on $G^*_2$, all being two dimensional. The generic one corresponds to the case case $k \neq 0, y \neq 0$ ; a second one corresponds to the case $k \neq 0, y = 0$ and a third one corresponds to the case $k = 0, y \neq 0$. The last one corresponds to the case $k=0, y=0$. In this paper we study in details the generic one and we summarize for the four orbits in a table just before the references.

3 **Yank and Hooke's Constant**

Let us define $q$ and $\tau$ by $f = kq, f = y\tau$ which means that $q = v\tau$ and $\tau = sq$ where $v = \frac{y}{k}$ is an invariant velocity and $s = \frac{1}{v}$ is an invariant slowness.

We then verify from (2.8) that the quantities

$$U = e - k\frac{q^2}{2} + pv \;,\; \pi = p - y\frac{\tau^2}{2} + es$$

are Ad*-invariants. Note that $U$ has energy as physical dimension while $\pi$ has that of a linear momentum. They are related by the relation

$$U = \pi v$$

Note also that quadratic form

$$\Psi = 2ke - f^2 + 2py$$

is $G_1$-invariant. It means that (2.8) realize the first central extension of the Aristotle Lie group as a nilpotent subgroup of the pseudo-orthogonal (semi-simple) Lie group SO(3,2).

We can study the "dynamics" on the generic orbit with respect to the time translations with respect to the space translations. If we choose time translation the orbit will be characterised beside $k$ and $y$, by the energy $U$. We will denote the orbit by $O_{(k,v,U)}$. If we choose space translation the orbit will be characterised, beside $k$ and $y$, by the momentum $\pi$. We will denote the obit by $O_{(k,s,\pi)}$.

3a. **Evolution with time**

The generic orbit is endowed with the symplectic 2-form $\sigma = dp \wedge dq$ and the symplectic realization of $G_1$ on is then

$$\Phi_{(x,t,\varsigma)}(p,q) = (p - ft - k\varsigma + y\frac{t^2}{2}, q + x - vt) \qquad (3.1)$$

Note that $q$ transforms like in Galilean kinematics [5]. To study the physics of the orbit we introduce the contact manifold $\Re x\, O_{(k,y,U)}$ and endow it with the 2-form

$$\sigma(t) = \sigma - dH \wedge dt \qquad (3.2)$$

where $H(p, q, t)$ is the hamiltonian function conjugated to time.

If $(p_0, q_0, 0)$ is the initial state of the system and if $(p(t), q(t), t) = \Phi_{(0,t,0)}(p_0, q_0, 0)$, we then have from (3.1) that

$$p(t) = p_0 - ft + y\frac{t^2}{2}, q(t) = q_0 - vt \qquad (3.3)$$

and that $\Phi_{(x,t,\varsigma)}(p_0,q_0,0) = (p(t)-k\varsigma, q(t)+x, t)$. We can interpret the component

$P(t) = -ft + y\dfrac{t^2}{2}$ of $p(t)$ as a potential momentum (impulse).

The Lie algebra $G_1$ is represented on this contact manifold by the hamiltonian vector fields

$$\Phi_{t^*}(E) = \dfrac{\partial}{\partial t} + -k(q-vt)\dfrac{\partial}{\partial p} - v\dfrac{\partial}{\partial q},\ \Phi_{t^*}(P) = \dfrac{\partial}{\partial q},\ \Phi_{t^*}(F) = -k\dfrac{\partial}{\partial p} \qquad (3.4)$$

These give rise to the equation of motion (Hamiltonian equations)

$$\dfrac{dq}{dt} = -v \qquad (3.5a)$$

$$\dfrac{dp}{dt} = -kq + c(t)\dfrac{dq}{dt}, \qquad (3.5b)$$

where $c(t) = kt$ is a damping coefficient. We see that the right hand side of (3.5b) is a contribution of a Hooke's force $kq$ and a damping force $c(t)\dfrac{dq}{dt}$. Finally the hamiltonian is

$$H = \dfrac{1}{2}kq^2 - (p + c(t)q)v \qquad (3.6)$$

where $c(t)q$ is a damping momentum.

3b. *Evolution with space*

This time we endow the orbit with the symplectic 2-form

$$\sigma = de \wedge d\tau \qquad (3.7)$$

and the symplectic realization of $G_1$ become

$$\Phi_{(x,t,\varsigma)}(e,\tau) = (e + fx + y(\varsigma - xt) + k\dfrac{x^2}{2}, \tau - t + sx) \qquad (3.8)$$

We then note that $\tau$ transforms as in the Carroll kinematics [5].

To study the physics of the orbit we introduce the contact manifold $\Re x\, O_{(k,s,\pi)}$ and endow it with the 2-form

$$\sigma(t) = \sigma + d\Pi \wedge dx \qquad (3.9)$$

where $\Pi$ has the physical dimension of a linear momentum. We then see that

$$\Phi_{(x,t,\varsigma)}(e_0, \tau_0, 0) = (e(x) + y\varsigma, \tau(x) - t, x) \qquad (3.10)$$

where

$$e(x) = e_0 + fx + k\frac{x^2}{2} \qquad (3.11a)$$

and

$$\tau(x) = \tau_0 + sx \qquad (3.11b)$$

Note that the component $fx + k\dfrac{x^2}{2}$ in (3.11a) is a potential energy and that $\tau(x)$ can be considered as a travel time like in ray tracing [10].

One can then verify from (3.10)-(3.11) that the Lie algebra $G_1$ is represented on the contact manifold by the hamiltonian vector fields

$$\Phi_{x^*}(E) = -\frac{\partial}{\partial \tau},\ \Phi_{x^*}(P) = \frac{\partial}{\partial x} + k(q+x)\frac{\partial}{\partial e} + s\frac{\partial}{\partial \tau},\ \Phi_{x^*}(F) = y\frac{\partial}{\partial e} \qquad (3.12)$$

The equations of evolution are then

$$\frac{d\tau}{dx} = s \qquad (3.13a)$$

$$\frac{de}{dx} = y\tau + W(x)\frac{d\tau}{dx} \qquad (3.13b)$$

where $W(x) = yx$ is a power. We then verify that the hamiltonian is

$$\Pi = \frac{1}{2}y\tau^2 - (e - W(x)\tau)s \qquad (3.17)$$

We recognise in the right hand side of (3.13b) a contribution of a force proportional to time with yank as the coefficient and a damping force proportional to a slownes, with a power as the coefficient .The right hand side of (3.13a) is a slowness which is used in ray tracing to compute the time travel of waves [10].

## 4. CONCLUSION

Finally we see that *if the physics of the orbit is studied with respect the evolution in time , it represents an elementary with internal energy U in a position-momentum (q,p) state under the conjugation of a Hooke's force kq and a damping force* $c(t)\frac{dq}{dt}$, $\frac{dq}{dt}$ *being a velocity as in particle mechanics.If the physics of the orbit is studied with respect the evolution in space , it represents an elementary with internal linear momentum Π in a energy-time (e, τ ) state under the conjugation of a kind of Hooke's force yτ and a damping force* $W(x)\frac{d\tau}{dx}$, $\frac{d\tau}{dx}$ *being a slowness as in time travel waves*.

*The results for all the orbits are summarised in the table below.*

## Summary of the $G_1$-orbits

| | Generic orbit $O_{(k,v,U)}$ (evolution with time) | Generic orbit $O_{(y,s,\pi)}$ (evolution with space) | $O_{(k,U)}$ | $O_{(y,U)}$ | $O_f$ |
|---|---|---|---|---|---|
| **invariants** | $k, v, U = e - k\frac{q^2}{2} + pv$ | $k, s, \pi = p - y\frac{\tau^2}{2} + es$ | $k, U = e - k\frac{q^2}{2}$ | $y, \pi = p - y\frac{\tau^2}{2}$ | $f$ |
| **Hamiltonian** | $H = k\frac{q^2}{2} - (p + c(t)q)v$ | $\Pi = y\frac{\tau^2}{2} - (e - W(x)\tau)s$ | $H = k\frac{q^2}{2}$ | $\Pi = y\frac{\tau^2}{2}$ | $H = fq$ |
| **evolution** | $p(t) = p_0 - ft + y\frac{t^2}{2}$, $q(t) = q_0 - vt$ v is a velocity | $e(x) = e_0 + fx + k\frac{x^2}{2}$, $\tau(x) = \tau_0 - sx$ s is a slowness | $p(t) = p_0 - ft$, $q(t) = q_0$ | $e(x) = e_0 + fx$, $\tau(x) = \tau_0$ | $p(t) = p_0 - ft$ $q(t) = q_0$ |
| **motion equations** | $\frac{dp}{dt} = -kq + c(t)\frac{dq}{dt}$, $\frac{dq}{dt} = -v$ | $\frac{de}{dx} = y\tau + W(x)\frac{d\tau}{dx}$, $\frac{d\tau}{dx} = s$ | $\frac{dp}{dt} = -kq$, $\frac{dq}{dt} = 0$ | $\frac{dp}{dt} = y\tau$, $\frac{dq}{dt} = 0$ | $\frac{dp}{dt} = -f$ $\frac{dq}{dt} = 0$ |

## 5.REFERENCES